\documentstyle[prl,aps,floats,twocolumn,graphics,epsfig]{revtex}

\title{Zero-temperature responses of a 3D spin glass in a field}

\author{F.~Krzakala$^{1}$,
        J. Houdayer$^{2}$,
        E.~Marinari$^{3}$,
        O.~C.~Martin$^{1}$ and
        G.~Parisi$^{3}$}

\address{
 $^{1}$ Laboratoire de Physique Th\'eorique et Mod\`eles Statistiques,
b\^at. 100, Universit\'e Paris-Sud, F--91405 Orsay, France.\\
 $^{2}$ Institut f\"ur Physik, D-55099,
and Max Planck Institut f\"ur Polymerforschung, D-55128 Mainz, Germany. \\
 $^{3}$ Dipartimento di Fisica, INFM, SMC and INFN, 
Universit\`a di Roma {\em La Sapienza}, P. A. Moro 2, 00185 Rome, Italy.\\
}
\date{\today}
\draft

\begin{document}
\twocolumn[ \hsize\textwidth\columnwidth\hsize\csname
@twocolumnfalse\endcsname 
\maketitle

\begin{abstract}
We probe the energy landscape of the 3D Edwards-Anderson spin glass
in a magnetic field to test for a spin glass ordering. We find
that the spin glass susceptibility is anomalously large on the lattice
sizes we can reach. Our data suggest that a transition from the
spin glass to the paramagnetic phase takes place at $B_c \approx 0.65$, 
though the possibility $B_c=0$ cannot be excluded. We also discuss the
question of the nature of the putative frozen phase.
\end{abstract}


]

Ising spin glasses~\cite{Young98} have been studied intensively for over two
decades; nevertheless, even the basic issue of the nature of the phase
diagram is still unsettled. There are two main schools of thought in the
long standing debate over whether a spin
glass phase can exist in the presence of a magnetic field $B$. 
In the mean field picture~\cite{MezardParisi87b} the spin glass 
phase exists up to some critical field value $B_{AT}(T)$, {\it i.e.}, up to 
the so called ``de Almeida-Thouless'' line~\cite{AlmeidaThouless78}
(AT) where replica symmetry breaking (RSB) arises. In the
droplet picture~\cite{BrayMoore84,FisherHuse86}, any 
non-zero magnetic field kills the spin glass ordering and makes the system
paramagnetic.

Experimental evidence in favor of each school of thought has been
claimed~\cite{ItoArugakatori94,MattssonJonsson95}, but no consensus
has emerged. On the computational side, the approaches using Monte
Carlo have been hindered by the large finite size effects present in
$d=3$~\cite{CaraccioloParisi90a,GrannanHetzel91,KawashimaIto93}.
Because of this, dynamical out-of-equilibrium approaches have been
used ({\it e.g.}, \cite{ParisiRicci98,MarinariParisi98d} for $d=4$ 
and \cite{MarinariParisi99,MarinariParisi00d} for $d=3$); those works hint
at the existence of a phase transition in field, but it
is preferable to have a direct test using a system {\it in}
equilibrium.  For such a test, one should stay away from the critical
point at $B=0$. To achieve this, we focus in this work on the $T=0$ line in
the $B$-$T$ phase diagram; then the spin glass ordering can be tested
through the behavior of excitations above the ground state.

We begin by discussing possible scenarios for the spin glass
ordering when $B\ne 0$. In our first approach, we extract low energy
excitations and investigate their size as a function of $L$ 
and of $B$; we also consider some of their topological
properties.  In a second approach, we compare ground states with periodic
and anti-periodic boundary conditions, and extract a ``magnetic
penetration length''.  We use extrapolations and finite size scaling
to estimate the critical field $B_c$ where the system becomes
paramagnetic. Our results suggest $B_c \approx 0.65$.

\paragraph*{The model and spin glass ordering ---}

We consider the EA Hamiltonian on a $3D$ $L^3$ periodic cubic lattice:
\begin{equation}
\label{eq_H_EA}
H_J(\{S_i\}) \equiv - \sum_{<ij>} J_{ij} S_i S_j - B \sum_{i} S_i\ .
\end{equation}
In this Edwards-Anderson (EA) model, 
the sum is over all nearest neighbor spin pairs, $S_i = \pm 1$, and
$B$ is the strength of the magnetic field. The quenched couplings
$J_{ij}$ are independent random variables taken from a Gaussian
distribution of zero mean and unit variance.  At $B=0$, it is widely
believed that the low $T$ behavior of this system is characterized by
a frozen but random ordering of the local magnetizations: $\langle S_i
\rangle \ne 0$, where $\langle . \rangle$ is the thermal average.
When $B>0$, there is no up-down symmetry and the fact that $\langle
S_i \rangle \ne 0$ is not relevant.  Instead, one relies on the
correlation length $\xi$: $\xi = \infty$ in the spin glass phase,
whereas $\xi$ is finite in the paramagnetic phase.

We study the system at $T=0$ and ask what signature
corresponds to a $\xi = \infty$ behavior of 
$T>0$ system.  Given the ground state in the presence of $B$,
consider droplet excitations, {\it i.e.}, excitations that are of
minimal energy at fixed ``size'' (taken generally as the ``radius''
$r$ of
the connected cluster of spins that are flipped).  Let $\theta$ be the
exponent~\cite{FisherHuse86} giving the characteristic excitation
energy of these droplets: $E \approx r^{\theta}$. If the
distribution of the rescaled energies, $P_d(E / r^{\theta})$, has a
non-zero density at zero argument, then there are enough thermally
activated droplets of all sizes to make the correlation length
infinite for low enough temperatures.  
We have three main scenarios:

(1) The ``mean field'' scenario: if $B$ stays below
some critical value $B_{AT}$ there is RSB~\cite{MezardParisi87b}; for
$B > B_{AT}$ the behavior is paramagnetic.  Because of RSB, spin-spin
correlation functions do not cluster when $B < B_{AT}$: any reasonable
definition of $\xi$ leads to $\xi = \infty$ below $B_{AT}$.

(2) The ``scaling/droplet'' scenario: as soon as $B>0$ the system is
paramagnetic. Motivation for this scenario comes from the
large field limit: there, $\theta = 3$ because excitation
energies grow {\it linearly} with the number of spins flipped. Based on
fixed-point arguments, one then expects $\theta = 3$ 
and $P_d(0)=0$ for all $B > 0$. Because of this 
last property, large scale droplets cannot be
thermally activated in the presence of a field (or equivalently when
there is an extensive magnetization).

(3) An ``intermediate'' scenario: the spin glass order survives at
positive $B$, but there is no RSB and thus no AT
line. $\theta(B)$ may be arbitrary while $B_c$ is defined
as the field
where $P_d(0)$ goes to $0$, separating the regime of $\xi =\infty$ 
from the one where $\xi<\infty$.  Since this scenario has
large scale thermally activated droplets {\it co-existing} with 
an extensive magnetization, we will refer to
it as the MAD (Magnetization And Droplets) scenario.

\paragraph*{Droplet sizes ---}
In our first approach we generate low energy clusters of spins
according to a {\it one spin flip} method.  Given the ground state
$C_0$ of $H$, we randomly choose a spin $S_{i_0}$ and force its
orientation to be opposite from what it is in the ground state. Then
we recompute the new ground state $C$ given this constraint.  (The
numerical algorithm for computing the ground states is described
in~\cite{HoudayerMartin01}.) The difference between $C$ and $C_0$ is a
connected cluster of spins whose volume we denote by $V$; it is the
lowest energy cluster among all those containing $S_{i_0}$.  The mean
cluster size $\langle V \rangle$ is analogous to the spin glass
susceptibility $\chi_{SG} = L^3 \langle (q - \langle q\rangle)^2
\rangle$, where $q$ is the standard spin overlap.  A diverging
$\langle V \rangle$ as $L \to \infty$ corresponds to a spin glass
phase, while a bounded $\langle V \rangle$ is the signature of a
paramagnetic phase.

\begin{figure}
\centerline{\hbox{\epsfig{figure=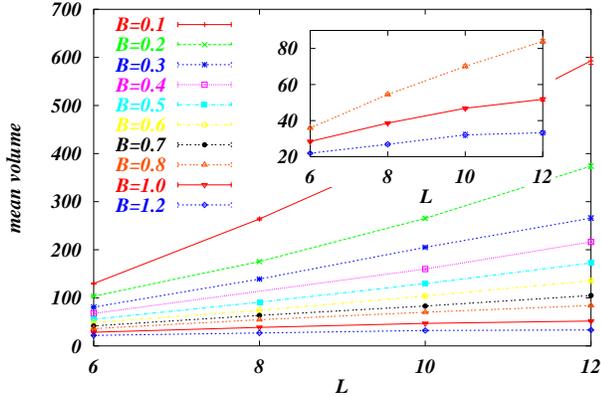,width=8cm}}}
\caption{Mean volume of the droplets extracted by the one-spin flip
method as a function of $L$. From top to bottom: $B=0.1$, $0.2$,
$\cdots$. In the insert $B=0.8$, $1.0$ and $1.2$.}
\label{fig_size}
\end{figure}

For each value of $B$ ($0.1 \le B \le 1.2$, in steps of $0.1$) we have
generated $3000$ disorder instances for $L=6,8,10$, and $1000$ for
$L=12$. In figure~\ref{fig_size} we display the mean volume $\langle V
\rangle$ of the droplets generated by this one-spin flip method.
(The spin $S_{i_0}$ selected for flipping
is chosen at random amongst the $25\%$ spins with the largest local
fields; this enhances the signal but does not affect our conclusions.)
At $B \ge 1$ we see signs of $\langle V \rangle$ saturating when $L$
grows: these large $B$ values are in the paramagnetic phase.  For
smaller values of $B$, $\langle V \rangle$ grows significantly with
increasing $L$.  In the droplet picture there is a magnetic length
$\ell_B$ which acts as a cut-off: $\langle V \rangle$ will grow as a
power of $L$ until $L \approx \ell_B$, and thereafter it will
saturate. Furthermore, this cut-off length grows as $B$ decreases,
$\ell_B \approx (\Upsilon / B)^{{\frac{3-2\theta}{2}}}$.  On the
contrary, in the MAD and mean field scenarios $\langle V \rangle$
diverges for all $B \le B_c$.

Consider for example the data at $B=1.0$. If we use the
estimates on the largest lattices \cite{CieplakBanavar90}
available to date ($\Upsilon\approx 1.78$ and $\theta \approx 0.19$), 
we have $\ell_B(B=1.0)\approx 2.13$. 
$\langle V \rangle$ should be roughly the cube of this number,
while we find $\langle V \rangle = 52$ at $L=12$.
Such a large value of $\langle V \rangle$ is unexpected in 
the droplet picture. Unfortunately,
since the values of $\Upsilon$ and $\theta$ have large uncertainties,
one cannot be more quantitative than that. In fact, the curves displayed in
figure~\ref{fig_size} could be interpreted either by saying that for
$B \le 0.8$, $\ell_B$ is comparable to our largest lattice sizes, or
that $\langle V \rangle$ diverges as $L \to \infty$ even when $B$ is
not too small, for instance for $B < 0.6$.  To further test the
different scenarios, we follow~\cite{HoudayerKrzakala00} and focus now
on the {\it topology} of our excitations.

\paragraph*{Fraction of sponge-like excitations ---}
In the mean field (RSB) picture, macroscopically distinct valleys
differing by energies of $O(1)$ arise with positive probability when
$B<B_{AT}$ and with zero probability for $B>B_{AT}$. (When the volume
is finite, this last probability should go to zero exponentially in
$L$.)  These valleys differ by excitations that span the whole system
and that {\it wind} around the lattice. We say that the cluster
associated with one of our excitations is {\it sponge-like} if both it
and its complement wind around all three directions $(x,y,z)$ of the
lattice. This motivates our measuring the fraction $f(B,L)$ of the
excitations generated by our one-spin flip method that are
sponge-like. (For low field values, we observe that the
flipped cluster sometimes contains more than $0.5 L^3$ lattice sites;
since such events disappear as $L \to \infty$ in all three scenarios,
we have excluded them from our analysis in order to reduce the finite size
corrections.)  Let $f^*(B)$ be the large $L$ limit of $f(B,L)$.
$f^*(B)$ is an order parameter for RSB: it is zero for $B
> B_{AT}$ and positive for $B<B_{AT}$.

\begin{figure}
\centerline{\hbox{\epsfig{figure=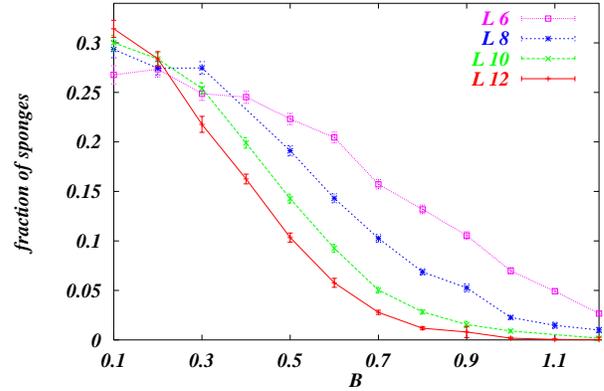,width=8cm}}}
\caption{Fraction of sponge-like excitations versus $B$.}
\label{fig_1sf_topo}
\end{figure}

In figure~\ref{fig_1sf_topo} we show our values for the fractions
$f(B,L)$. As expected at large $B$, $f$ goes to zero with
increasing $L$.  If there is RSB, the large $L$ curves will converge
to a limiting function that intersects the $x$ axis at $B_{AT}>0$. On
the contrary, in the other pictures where replica symmetry is not
broken, the limiting curve is $f^*(B)=0$ for all $B>0$.

What can be said from these data within the RSB scenario?  The order
parameter should behave as $f^*(B) \approx (B_{AT}-B)^{\beta}$ with
$\beta$ less or equal to its mean field value of $1$. Thus
we expect $f^*$ to be convex for $B$ not too close
to $0$. Now if we make the reasonable assumption that
$f(B,L)$ converges to $f^*$ monotonically, 
then we get a bound on $B_{AT}$ by taking the tangent to the
$L=12$ curve; this leads to $B_{AT} \le 0.65$. Of course, we can
be more realistic and say that the curves for different $L$ 
drift towards $f^*$ in a smooth way. If we do that and
extrapolate the curves by eye, we find $B_{AT} \le 0.4$.
Obtaining a less subjective estimate requires using
finite size scaling and parametrizing $f^*$, but 
our small range in $L$ makes such an attempt futile.

Consider now the data from the point of view of the two scenarios with
no RSB. We have performed the finite size scalings $f(B,L) = L^{-
\beta / \nu } F \lbrack (B_c - B) L^{1 / \nu} \rbrack$ where $B_c=0$
in the scaling/droplet scenario and $B_c>0$ in the MAD scenario. For
each putative value of $B_c$, we first adjust $\beta/\nu$ so that $F$ is
nearly $L$ independent at $B=B_c$, and then we adjust
$\nu$ for the curves to superpose as well as possible. For both
$B_c>0$ and $B_c=0$ the data collapse is far from perfect, so we
cannot use the chi squared for a quantitative analysis.  However, the
quality of the superposition can be judged by eye. For $B_c \approx
0$, the superposition is reasonably good and leads to $\nu=1.25$ and
$\beta \approx 0$ (recall that sponges seem to arise with a finite
probability in zero field). As $B_c$ is increased, the superposition
first becomes less good but then becomes better again at intermediate
values with a local optimum at $B_c \approx 0.65$; there we find 
$\nu \approx 0.8$ and
$\beta \approx 2.0$.  Without a reliable parameterization of the finite size
effects and larger lattice sizes, it is not possible to go
beyond these qualitative conclusions.

\paragraph*{Periodic-anti-periodic computation ---}
In our second approach we perturb the couplings $J_{ij}$ and probe how
this affects the ground state.  First we compute the
ground state with periodic boundary conditions. Then we change the
sign of all the $J$s cutting a given vertical plane $\Pi$; this
amounts to applying anti-periodic boundary conditions across this
plane. Finally we re-compute the ground state for this new system. In
contrast to the zero field case, gauge invariance is broken here and
the plane $\Pi$ where the $J$s are reversed is measurable. Our reason
for choosing this particular perturbation is that it is
translationally invariant in two directions; this leads to good
statistics on the observables.

Given this ``perturbation'', how does the difference between the two ground
states look?  At high fields, a few spins will be flipped in the
immediate neighborhood of the plane $\Pi$. As $B$ decreases, the
region affected by the perturbation will broaden; one can introduce a
``magnetic penetration length'' $\ell_P$ as a measure of the region's
width. In the paramagnetic phase, $\ell_P$ is finite, while it is
plausible that it grows with the system size $L$ in a spin glass
phase.  It is possible that in a droplet picture $\ell_P = \ell_B$,
but we do not develop this issue here.

\begin{figure}
\centerline{\hbox{\epsfig{figure=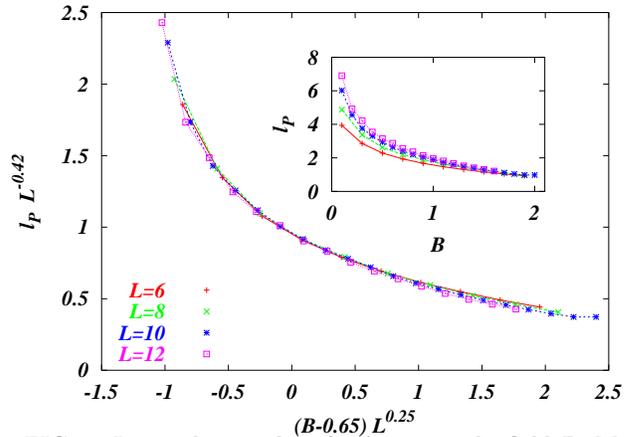,width=8.5cm}}}
\caption{Inset: the raw data for $\ell_P$
versus the field $B$. Main plot: $\ell_P$ rescaled by $L^{0.42}$
versus rescaled distance from $B_c$.}
\label{fig_ell_P}
\end{figure}

To define $\ell_P$ from the perturbed region, we 
measure the spin-spin overlap of the two ground states and then
average in each plane parallel to $\Pi$. Let $Q(d)$ be this overlap,
with $d$ the distance from $\Pi$. In the paramagnetic phase $Q(d)$
will go to $1$ at large $d$, and the approach to this asymptote should
go as ${\rm exp} (-d/{\ell_P} )$. If on the contrary $\ell_P=\infty$, the
asymptote may be different from $1$, and the approach to that value
can be a power law in $d$. For each $L$ and $B$ we extract $\ell_B$ by
fitting $Q(d)$ to the form $1 + A \exp(-d/\ell_B)$. Above $B_c$, this
should lead to the measurement of the true $\ell_B$, while below $B_c$
the extracted $\ell_B$ will be an effective length that will grow with
$L$. (In these measurements, we used $40000$ disorder samples
at $L=6$ and $20000$ at $L=8$ for each $B$; for $L=10$, we used $5000$ samples
at the smallest $B$ value and up to $12800$ on the largest one, while
we used from $1000$ to $7000$ samples for $L=12$.)  The fits performed
using all the distances gave results very similar to the ones using
only the four (symmetrized) data points at the largest values of $d$.

We show the results of our best fits in the inset of
figure~\ref{fig_ell_P}. $\ell_P$ saturates at high fields, while it
increases with $L$ at low field. We also show in the figure a
finite size scaling plot in which we have set $B_c=0.65$;
$\ell_P$ has been rescaled by $L^{-0.42}$ and $(B-B_c)$ by
$L^{0.25}$. The data collapse is good, but similarly good results are
obtained for all values of $B_c<1.0$; the problem is that when
data do not have a big dynamic range, finite
size scaling can almost always be made to work. Thus for this
method we have no precision on the estimate of $B_c$.

We have also used other methods to estimate $\ell_P$ from the set of
spins that are flipped in the periodic-anti-periodic transformation;
these include the set's mean distance to $\Pi$, the mean extension of
its interface, etc...  All these measures lead to $\ell_P$ comparable
to $L$ when $B \le 0.7$ for our range of $L$s.  It is thus difficult
for us to say whether $\ell_P$ diverges when $B$ reaches $B_c>0$ or
only when $B \to 0$, but this is not unexpected, and has been a
serious shortcoming in all attempts to detect a spin glass ordering in
a field. Because of this, we now again appeal to topology: we measure
the proportion of instances where the flipped spins form a sponge-like
cluster.

\begin{figure}
\centerline{\hbox{\epsfig{figure=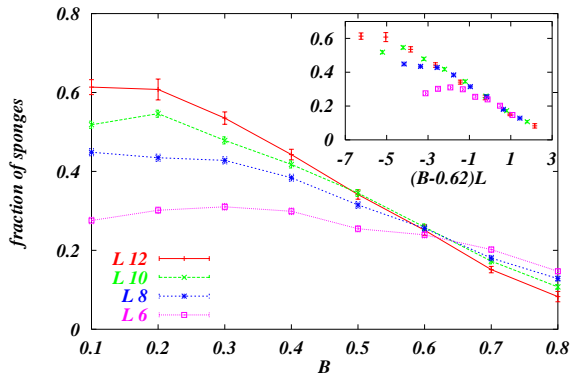,width=7.5cm}}}
\caption{Fraction of {\it sponge-like} clusters in 
the periodic anti-periodic approach. The insert gives 
the best finite size scaling fit ($B_c=0.62$, ${\nu}=1.0$).}
\label{fig_pap}
\end{figure}

The raw data for the fraction of sponges are plotted in
figure~\ref{fig_pap}. We observe a crossing point, suggesting a first
order jump in this fraction at $B_c \approx 0.65$.  This is probably
our clearest evidence for the existence of a spin glass phase with
$B_c>0$.  To perform finite size scaling, we take the
fraction of sponge-like events to be a function of
$(B-B_c)L^{1/{\nu}}$; the best data collapse is obtained for
$B_c \approx 0.62$ and $\nu \approx 1.0$ and is displayed in the figure's
insert. (Note that the data at low $B$ goes to the limiting
curve only slowly when $L$ increases.) These data are compatible
with a $B_c \approx 0.65$; however, since
the crossing points drift a bit,
a smaller value for $B_c$ cannot be ruled out.

\paragraph*{Discussion ---}
We have introduced and exploited a new approach for testing whether a
spin glass ordering arises in the presence of a magnetic field. The
major advantage of our method compared to Monte
Carlo is that since we work at $T=0$, we are far from 
the $B=0$ critical point, $T_c\approx 0.95$.  
Our main conclusion is that very plausibly in three dimensions,
a spin glass phase survives up to a critical
field $B_c \approx 0.65$; beyond that, the system becomes 
paramagnetic.

Our proposed value
is several times smaller than the mean
field value~\cite{HoudayerMartin99a}, $B_{AT}^{MF} \approx 2.1$.
Such a low value for $B_c$ makes it difficult to establish
without ambiguity that $B_c>0$. This problem is analogous to the
difficulty of showing that $T_c>0$ in $d=3$ spin glasses; to give
plausible evidence, finite size effects must be under excellent
control. Clearly, that is not yet the case here,
but the fact that our different approaches lead to consistent values
of $B_c$ gives some strength to our findings.

Finally, there is the question of a possible replica
symmetry breaking transition at $B_{AT}>0$.
Our numerical study leads us to suggest that $B_{AT} \le 0.4$
if it is indeed positive. This very low value is 
compatible with the findings
of~\cite{MarinariParisi99}. However, confirming such a value
via equilibrium measurements will require larger
lattice sizes than we can handle at present.

\paragraph*{Acknowledgments ---}

We thank M. M\'ezard, E. Vincent and F. Zuliani for stimulating discussions.
F.K. acknowledges financial support from the MENRT and
J.H. from the Max Planck Institute f\"ur Polymerforschung. E.M.
acknowledges an IUF
visiting professorship at Orsay during the early part of this work
and a travel grant from the ESF SPHINX program.  E.M. and
G.P. acknowledge support from the INFM {\em Center for Statistical
Mechanics and Complexity} (SMC).  The LPTMS is an Unit\'e de Recherche
de l'Universit\'e Paris~XI associ\'ee au CNRS.

\bibliographystyle{prsty}
\bibliography{../../../Bib/references}

\end{document}